# The Changing Geopolitics in the Arab World: Implications of the 2017 Gulf Crisis for Business

Jamal Bouoiyour
IRMAPE, ESC Pau Business school, France.
CATT, University of Pau, France.
E-mail: jamal.bouoiyour@uni-pau.fr

Refk Selmi
IRMAPE, ESC Pau Business school, France.
CATT, University of Pau, France.
E-mail: s.refk@yahoo.fr

**Abstract:** The international community was caught by surprise on 5 June 2017 when Saudi Arabia, the United Arab Emirates (UAE), Bahrain and Egypt severed diplomatic ties with Qatar, accusing it of destabilizing the region. More than one year after this diplomatic rift, several questions remain unaddressed. This study focuses on the regional business costs of the year-long blockade on Qatar. We split the sample to compare the stock market performances of Qatar and its Middle Eastern neighbors before and after the Saudi-led Qatar boycott. We focus our attention on the conditional volatility process of stock market returns and risks related to financial interconnectedness. We show that the Gulf crisis had the most adverse impact on Qatar together with Saudi Arabia and the UAE. Although not to the same degree as these three countries, Bahrain and Egypt were also harmfully affected. But shocks to the volatility process tend to have short-lasting effects. Moreover, the total volatility spillovers to and from others increase but moderately after the blockade. Overall, the quartet lobbying efforts did not achieve the intended result. Our findings underscore Qatar's economic vulnerability but also the successful resilience strategy of this tiny state. The coordinated diplomatic efforts of Qatar have been able to fight the economic and political embargo.

**Keywords:** 2017 Gulf crisis; stock markets; volatility; risk spillovers.

**JEL Classification:** F30, F36, F65, G11, G15.

# 1. Introduction

The policy disagreements at the center of the rift between Qatar and its Gulf neighbors seem not to be new. The anti-Qatar bloc has long regarded Qatar as too friendly to Iran, too annoying in its backing of Al Jazeera media network, without ignoring its perceived role in promoting the Muslim Brotherhood. Even though a variety of issues have been raised against Qatar, the most potent has been the United Arab Emirates (UAE) and Saudi Arabia's strong feeling of annoyance over avowed Qatar's support for Islamist movements. In addition, the competition between Qatar and the UAE for the leadership as the region's biggest financial hub. These developments put pressure on the Gulf region as an enduring political and security alliance, which became tangible in a diplomatic crisis that happened in 2014. At that time, Saudi Arabia, the UAE and Bahrain withdrew their ambassadors to Qatar since Doha did not put into effect a security agreement about non-interference in the internal affairs of the other GCC states. Nevertheless, their contemporary infighting exists in another dimension and might prompt a strategic shift in how the world looks at the geopolitics of the GCC. Indeed, the collective decision by Saudi Arabia, the UAE, Bahrain and Egypt to cut diplomatic and economic ties with Qatar on 5 June 2017 with a green light from President Donald Trump, has rattled nerves sending shockwaves around the world. These unprecedented tensions have exacerbated the uncertainty over the economic consequences of this crisis. The Qatari stock market lost about 10 percent in market value over the first four weeks of the boycott. Other GCC stock markets also fell in response to the blockade, though with varying extent. The Qatar's blockade disrupted supply chains, harmed the flow of goods and services, and provoked anxiety amongst many Gulf firms. Many businesses feared that escalating tensions could have serious consequences on business deals across the whole region.

Although there are a number of significant papers exploring the effects of economic, macroeconomic and financial uncertainty on asset price dynamics (Antonakakis et al. 2016;

Balcilar et al. 2016; Beckmann et al. 2017; Bouoiyour et al. 2018, etc.), rather less attention has been paid to the geopolitical risk and its impacts on international business. The vast majority of these studies indicate that the unstable political scene can have a pronounced impact on stock markets, portfolio allocation and diversification opportunities. The political turmoil exerts a significant influence on economic performance and asset prices (for instance, Guidolin and La Ferrara 2010). Likewise, geopolitical frictions and tensions lead to highest levels of uncertainty and prompt an ineffaceable mark on global markets (for example, Schneider and Troeger 2006; Zussman et al. 2008; Choudhry 2010). Conditional upon the type of the event, the effect of geopolitical uncertainty can be short-lived, have longer lasting impacts or yielding to shifts in markets affecting portfolio allocation and diversification decisions (inter alia: Pástor and Veronesi 2013; Kollias et al. 2013; Aslam and Kang 2015; Omar et al. 2016).

The escalated diplomatic tensions between Qatar and its Middle Eastern neighbors may cost them billions of dollars by slowing trade, investment and economic growth as it struggles with oil price collapse. During these times of distress, international investors and portfolio managers get poked and start to question the efficacy of their investment strategies. In fact, the most immediate impacts on businesses were debated intensively since the announcement of Qatar's isolation and it is still being debated, particularly because there is no sign of a resolution in sight to a diplomatic row between Qatar and its neighbors. Given these considerations, the present research seeks to investigate the business consequences of 2017 Gulf crisis. While there is no accordance on the relative importance of the costs of this crisis, we may get a feel of the possible boycott consequences by (1) comparing the conditional volatility process of the stock markets of Qatar and the boycotting countries before and after the blockade; and (2) testing whether this Gulf crisis has exacerbated the risk spillovers across the region. Stock markets may move together at times when traders and investors do not want them to (in particular, in times of heightened uncertainty) thus limiting the opportunities of portfolio diversification. As the portfolio risk control is a vital part of investment management, an accurate assessment of stock market volatility spillovers during the recent Qatar crisis would help investors to seek the best possible strategy to effectively manage volatility that is lowering portfolio returns.

Various GARCH (Generalized Autoregressive Conditionally Heteroscedastic) extensions are used to measure the volatility of stock markets before and after the Qatar's isolation. In general, the GARCH-type modeling allows depicting financial markets in which volatility can change, becoming extreme during periods of distress or sudden events and low during relatively calm periods. A simple regression model does not account for this variation in volatility exhibited in financial markets. GARCH processes differ substantially from homoskedastic econometric techniques, which suppose constant volatility and are utilized in basic ordinary least squares . The latter consists of lessening the deviations between data points and a regression line to fit those points. With asset returns, volatility is likely to vary significantly over specific time- periods and depend on past variance, making a homoskedastic model not optimal. GARCH models, being autoregressive, are conditional upon past squared observations and past variances to model for current variance. GARCH processes are widely employed in finance owing to their abilities to reduce errors in forecasting by controlling for errors in prior forecasting and, in turn, improving the accuracy of evolving predictions. Moreover, this study investigates the stock market volatility spillovers among Qatar, Saudi Arabia, UAE, Bahrain and Egypt by performing the forecast-error variance decomposition framework of a generalized Vector Autoregressive (VAR) model proposed by Diebold and Yilmaz (2012). The assessment of the interconnectedness of stock markets is of paramount importance for the understanding of a crisis and its propagation mechanism. Spillover effects in equity markets have been extensively evaluated in the extant literature (for example, Diebold and Yilmaz 2009; Engle et al. 2013). Throughout this study, we focus on the stock market volatility spillovers among Qatar and the boycotting countries while considering the uncertainty surrounding the Qatar diplomatic crisis. This method enables to assess the direction of spillover effects between various markets in an effort to identify the net transmitters or the net receivers of risk spillovers. To the best of our knowledge, it remains underexplored in recent empirical research. Such analyses would be useful for both portfolio risk managers and designers of policies aimed at safeguarding against increased political uncertainty surrounding the 2017 Qatar-GCC crisis.

Our findings reveal that the economic implications of the Qatar's isolation are likely to be costly to Qatar, Saudi Arabia and the UAE. For Bahrain and Egypt, the effect appears limited so far. After the blockade, the equities of Qatar, Saudi Arabia and UAE become more volatile and relatively more responsive to bad news. However, this volatility does not persist. Besides, our results suggest that the uncertainty surrounding the 2017 Gulf crisis increase, even partially, the volatility spillovers across Qatar, GCC and Egyptian stock markets. In short, our results suggest that the boycott did not achieve the expected outcome. The fact that the three main protagonists (i.e., Qatar, Saudi Arabia and the UAE) reacted in the same way to this crisis can be interpreted as a sort of victory for Qatar. The latter has shown resilience and a rapid and efficient adaptation. We advance throughout this research the main causes of this blockade and the strategy put in place by this tiny state to resist to Saudi and Emirati dominance.

The remainder of the study is organized as follows. Section 2 provides some insights about how the 2017 Gulf crisis started. Section 3 describes the methodology and the data. Section 4 reports and discusses the main empirical results. Section 5 concludes the paper and provides some economic implications of the Qatar diplomatic crisis.

## 2. Qatar-Gulf crisis : What we need to know ?

### *2.1. Saudi Arabia's dream of becoming the dominant Arab and Muslim power*

Saudi Arabia appears as the greatest regional power, because of its massive oil wealth, and also because of its new ambitions. The policy of wide-scale public works implemented by the government as well as foreign direct investment and banking and financial soundness have enabled Saudi Arabia to become the number one regional economy. Nevertheless, the economy of Saudi Arabia is entirely based on oil. The drop in oil prices since June 2014 created a certain obsession among Saudis with economic and political decline. Today, gigantic waves of change are sweeping across the Middle East region. The appointment of Prince Mohamed bin Salman (or MBS, as he is commonly referred to) as Crown Prince is part of this strategy. Previously it required the consent of the king's brothers and half-brothers of the king to pass on a project. Today, efficiency prevails. One should remember that the tradition in Saudi Arabia consisted of passing the 'Royal Scepter' among the sons of the kingdom founder, Ibn Saud, and not from father to son. This was a part of the internal politics driven by Ibn Saud many wives and dozens of children. When Saudi Arabia's king Abdullah bin Abdul-Aziz died in January 2015 at the age of 90, the candidates for his replacement were no longer young men. Nevertheless, the transfer of the role to the next generation intensified anxiety of an internal civil war breaking out between many princes, a war that might have damaged the existence of the House of Saud. To deal with increasing fears, the successor was his half-brother Salman who enjoyed the entire confidence of the other brothers. When the brother designated as Crown Prince was very old (about 80) and with failing health, royal decisions would be lengthy preventing the system from functioning effectively. Hence the mini-revolution that happened this year with the appointment of Prince MBS as Crown Prince. MBS was the sixth brother. Two main objectives are clearly identified. On the one hand, the achievement of a diversified economy and on the other hand, the ambition to embody the Sunni world, while associating Prince Mohamed Ben Zayed, the strong man of Abu Dhabi. MBS is taking the example of Abu Dhabi to develop its economy (Lavergne, 2018).

The tiny oil- and gas-rich Gulf state of Qatar has been a forerunner in this way. Indeed, during the last two decades or so, Qatar became one of the most influential countries of the Persian Gulf region and the Middle East. For a country established only in 1971 and with one of the smallest geographic and demographic sizes in the Middle East, Qatar became a surprising powerbroker dominantly owing to its financial muscle to project power and influence across the Middle East and North Africa region. Since the start of the Arab Spring in late 2010, the regional landscape has changed, and so has Qatar's policies. During the Arab Spring, Qatar moved away from its traditional foreign policy role as diplomatic mediator to embrace change in the Middle East and North Africa and to take an interventionist role as a leading supporter of the protest movements in the Middle East and North Africa. It is therefore not surprising to believe that the challenge launched in Qatar by Crown Prince MBS, along with three other countries in the region –Bahrain, Egypt[1] and UAE– is aimed at restoring the threatened supremacy of Saudi Arabia on the Arab and Muslim world and restore the strategic partnership of the United States with Saudi Arabia. In other words, the blockade imposed against Qatar by Saudi Arabia is not a matter of chance, but enters into a logic of Sunni world domination. The Qatar's challenge to Saudi Arabia is exacerbated by the fact that it adheres to Wahhabi creed. More accurately, Qatar's alternative adaptation of Wahhabism coupled with a long-standing links with the Muslim Brotherhood, make its relationship with Saudi Arabia more complicated and upraise it to a serious threat. The appointment of Prince MBS therefore has a dual purpose: economic efficiency and supremacy (Lavergne, 2018).

---

[1] Well prior to the blockade against Qatar, Egypt was a primary battleground for GCC countries striving for international influence. Even though Qatar backed the Muslim Brotherhood, SaudiArabia and the UAE supported the military regime of President Abdel Fatah al-Sissi. This explains to some extent how Egypt wound up in the center of a Gulf Cooperation Council conflicts with Qatar.

We realize, therefore, that the boycott hides a more insidious rivalry between Saudi Arabia and Qatar. To this 'inter-Sunni' rivalry one can add the rivalry between Saudi Arabia (Sunni) and Iran (Shiite). After the Geneva Agreements imposing strict controls on Iran's most sensitive nuclear work, Iran offered many opportunities to the Western business communities. This county has economic potential: Iran has an educated, urbanized and tech-astute population. It has a literacy rate of over 95 percent. The Yemen War should show the world, but especially the Western countries, the capacity of Saudi Arabia to defend its interests of the 'free' world, threatened by Iranian Shiite power. Likewise, this operation should assert the supremacy of the Wahhabi kingdom by bringing together a coalition of Arab-Muslim "friends". This show of force (in particular, boycott against Qatar and the Houthi offensive) may serve as a powerful signal given to the other partners of the Gulf Cooperation Council (GCC), reminding them of the Saudi leadership in the Middle East region.

### 2.2. David vs. Goliath? A misleading asymmetry

With a population of 250,000 and a surface area of 11,586 km2, it can be claimed that Qatar is a dwarf compared to Saudi Arabia (a population of 33 million and a surface area of 2,253,690 km2, according to the World Bank collection of development indicators). By shutting down all land, sea, and air crossings with the tiny energy-rich nation, the Saudi-led quarter anticipated that the surrender of Qatar is only a matter of days. The reality, however, is much different. As a small, vulnerable country situated in an unstable Middle Eastern politics, Qatar faces several challenges. Nevertheless, the tiny Qatar has used income from its wide gas reserves to bankroll its ambitious plans. Regardless of its size, it has played a significant leadership role, with a remarkable power in the Arab world. Qatar is also classified by the United Nations as the country with the highest human development among the Arab states. Also and in an attempt to prevent the damage from neighboring disputes, Qatar has often tried to strengthen its diplomatic relationship with multiple regional and international actors, by presenting itself as a friendly and helpful player. One cannot ignore the role of mediation[2] in branding Qatar's image on a political level.

It must be emphasized here that the Saudis, Emiraties and Qataris have familial relationships, implying that long-running family rivalries may be considered as one of the causes behind the big political issues. This may explain, to some extent, why the ongoing Qatar crisis poses a major dilemma for Kuwait and Oman. These two Arab Gulf states share the same interests in terms of preventing the Qatar crisis from prolonging. As competition of dominance intensifies, Officials in Kuwait City and Muscat are wary, as much as Qatar, about the Saudi leadership, exacerbated by Mohammed bin Salman's rise to power. Rather than following Saudi Arabia and its allies, Kuwait and Oman stayed neutral. The neutrality of these two Gulf countries provided leverage for Qatar, albeit without direct support. But it must be mentionned at this stage that Kuwait appears as the main mediator among the warring parties, and Oman endorsed diplomacy while enhancing its links with Qatar. Beyond the reforms undertaken by the Qatari authorities to deal with the crisis, there have been other reasons why the impact of the blockade imposed aganist Qatar has not been as hurtful as it might have been. Among the potential reasons, one can cite the Omani and Kuwaiti foreign policy strategies. Even though Saudi Arabia, the UAE and Bahrain have imposed their trade and investment boycott against Qatar, Oman and Kuwait have chosen to stay resolutely above the fray.

To this we must add the role played by the US in this region. The US president Donald Trump accused Qatar in June 2017 of funding terrorism. Then and while attempting to change Trump's mind about Qatar, the emir of Qatar has spent millions of dollars hiring lobbyists and powerful American brokers to Doha. A few months later, Trump thanked Qatar for its efforts to combat terrorism and extremism in all forms in an apparent contradiction of previous statements. It must be stressed, nevertheless, that Saudi Arabia and the UAE have been indispensable backers of Egypt since the overthrow of Mohamed Morsi. In an attempt to support the Egypt's president Abdel Fattah al-Sisi, Saudi Arabia and the UAE have employed a number of financial tools, such as deposits into the Egyptian Central Bank, donations of oil and gas shipments, and promises of increased foreign direct investments in different sectors. In short, Saudi Arabia and the UAE has emerged as the leading supporters of Egypt's military rulers. As for Bahrain, it has lost all autonomy since the Saudi-led intervention on mid-March 2011 to assist the Bahraini government in subduing an anti-government protests in the country. This multiplication of actors does not stop there. The Turkish president has been a major supporter of Doha since the quartet cutties and imposed boycott against Qatar . Also and according to Qatar's Chamber, Turkey is one of Qatar's major customers for non-oil exports. Likewise, Qatar's pledge of aid to Turkey has strengthened the two countries' alliance. In August 2018, the emir of Qatar pledged to invest 15 billion dollars in Turkey, which grapples with a currency crisis that made the lira collapse by about 45 percent against the US dollar. In the same order of ideas, Doha sees its links with Tehran as vital to its economic and security interests. Iran's President Hassan Rouhani also announced his country's support of Doha during this crisis. Qatar's ties to Iran are of paramount importance to guard its natural resources, as the countries share the biggest gas field in the world. Since the blockade, Iran and Qatar ties have improved. As a response to the 2017 Gulf crisis, Iran voiced its support for the Qatari government, consolidating its alliance with the small Gulf state. Iran's trade with Doha totaled 250 million dollars in2018, registering a sharp rise of 2.5 percent compared with 2017.

---

[2] One of the major factors in changing the way in which Qatar is viewed regionally and globally is the creation of the Al-Jazeera Channel.

In sum, the escalated tension between Qatar and the quartet is in many ways a friction about the exercise of economic foreign affairs. Qatar utilizes its economic resources to support Muslim Brotherhood, and Saudi Arabia and the UAE see this support as extremely threatening to their own regimes. These competing visions have continuously tried to achieve their regional dominance by reinforcing aid and investment patterns which have the potential to contort the political economy of the whole region. By means of relatively new econometric techniques, we will see throughout the rest of our study, the consequences of this stunning political development on the subject of interest, in particular whether an escalating Gulf geopolitical crisis has intensified the market volatility in the region. More globally, this study seeks to identify the winners and the losers of Qatar standoff. It is important to remember that Qatar has always been aware of its vulnerability and has managed its business with dexterity (multiplying foreign partners, strengthening the management of gas resources, and pursuing investment mediation) despite the economy's reliance on the hydrocarbon sector. Certainly, this tiny state is confronted with several challenges due to the diversity of its population as well as its transformation from a traditional society to a modern state, with all that may involve in terms of changing societal and cultural norms. All this underscores the complicacy of the analysis of this region and the intricacy of the interests of several powers, without overlooking the fact that this region is the holder of the largest oil reserves in the world. Given all that, the match between the two protagonists (i.e., Saudi Arabia and Qatar) is not between unbalanced forces as one might think. Qatar is not fully isolated and Saudi Arabia is not as powerful as the statistics might suggest. This can be advanced as an element of explanation for Qatar's resilience of the blockade imposed by Saudi Arabia and its allies.

3. **Methodology and data**

This study performs a variety of econometric methods (a) to answer what Qatar diplomatic crisis means for the stock market performances of Qatar, Saudi Arabia, the UAE, Bahrain and Egypt, and (b) to explore the stock market volatility interdependence between Qatar and the boycotting countries before and after the 2017 Gulf crisis.

### 3.1. Measuring volatility using GARCH-type modeling

Although it seems not easier to quantify the full costs of 2017 Gulf crisis, the present research uses relatively new techniques in an attempt to provide fresh insights that may help policymakers to make the best possible decisions to deal with uncertain exposure. Given the challenges in consistently capturing the dynamic relationship between geopolitical uncertainty and stock markets, this paper seeks to compare the stock market volatility of Qatar and the boycotting countries before and after the blockade. There is a wide-spread perception in the financial press that volatility of asset returns has been changing markedly. The standard models consider that the distribution of asset returns is stable, implying that economic agents formulate their expectations at the same way over time. This evidence is far from reality, since during periods of great agitation (i.e., adverse changes, crisis, political tensions and sudden shocks, etc.), the variance-covariance of returns may move excessively. As a result, the standard techniques are unable to properly capture the conditional volatility process and to account for transitory and permanent components, shifts possibly stemming in the investigated variables. It is therefore relevant to examine the validity of this perception and to determine the features of changing volatility dynamics. Table A.1. (Appendix) succinctly reviews different GARCH models that account for various features (asymmetry, nonlinearity, regime shifts, etc.) that may be embedded in data. Since no single measure of volatility has dominated the existing empirical literature, the appropriate model able to properly depict the volatility of stock indices for Qatar and the boycotting countries is selected throughout this study using the Akaike information criterion (AIC). The latter helps to judge the quality of conditional variance estimation in terms in terms of trade-off between goodness of fit and model parsimony.

### 3.2. Measuring the volatility spillover effects

After evaluating the changing volatile behaviors of Qatar, Saudi Arabia, UAE, Bahrain and Egypt stock markets to the 2017 Gulf crisis, we now concentrate on the impact of this diplomatic crisis on the extent of volatility transmission across these countries. This work does not focus on the effect over the day relative to the boycott announcement only; rather it assesses the spillover effects before and after the decision of blockade on Qatar.[3] To this end, we include the conditional volatility series[4] to a generalized VAR framework (Diebold and Yilmaz, 2012). The conducted volatility transmission analysis covers three aspects.

First, we determine the total volatility spillover index which measures what proportion of the volatility forecast error variances comes from spillovers. Let:

$$x_t = \phi x_{t-1} + \varepsilon_t \quad (1)$$

where $x_t = (x_{1,t}, x_{2,t})$ and $\phi$ is a 2*2 parameter matrix; $x$ will be considered as a vector of the considered stock volatilities.

By covariance stationarity, the moving average representation of the VAR is denoted:

$$x_t = \Theta(L)\varepsilon_t \quad (2)$$

where $\Theta(L) = (I - \phi L)^{-1}$

Second, we consider 1-step-ahead forecasting. The optimal forecast is given by:

$$x_{t+1,t} = \phi x_t \quad (3)$$

with corresponding 1-step-ahead error vector:

$$e_{t+1} = x_{t+1} - x_{t+1,t} = A_0 \mu_{t+1} = \begin{bmatrix} a_{0,11} & a_{0,12} \\ a_{0,21} & a_{0,22} \end{bmatrix} \begin{bmatrix} \mu_{1,t+1} \\ \mu_{2,t+1} \end{bmatrix} \quad (4)$$

---

[3] We test whether the volatility spillovers among Qatar, Bahrain, Egypt, Saudi Arabia and UAE stock returns has been exacerbated over the period witnessing heightened uncertainty over the 2017 Gulf crisis.
[4] The conditional volatility of each stock index is determined through the best GARCH model chosen using the the Akaike information criterion.

In particular, the variance of the 1-step-ahead error in forecasting $x_{1,t}$ is $a_{0,11}^2 + a_{0,12}^2$, and the variance of the 1-step-ahead error in forecasting $x_{2,t}$ is $a_{0,21}^2 + a_{0,22}^2$. There exist two possible spillovers in our example: $x_{1t}$ shocks that exert influence on the forecast error variance of $x_{2t}$ (with contribution $a_{0,21}^2$), and $x_{2t}$ shocks that affect the forecast error variance of $x_{1t}$ (with contribution $a_{0,12}^2$). Hence the total spillover effect is equal to $a_{0,12}^2 + a_{0,21}^2$. Having outlined the Spillover Index in a first-order two-variable VAR, it is easier to generalize this to a dynamic framework for a p$^{th}$-order N-variable case.

Third, we quantify the net directional volatility spillovers for stock indices, in order to identify which of the considered countries are net volatility importers, and which of them are stress volatility exporters. At this stage, we decompose the total spillover index for stock volatilities into all of the forecast error variance components for variable *i* coming from shocks to variable *j*, for all *i* and *j*.

### 3.3. Data and descriptive statistics

The data of Qatar, Bahrain, Egypt, Saudi Arabia and UAE stock price indices were collected from DataStream (Thomson Reuters). To evaluate the business costs of Qatar diplomatic crisis on Qatar and its neighbors, we compare two equal periods prior to and post the blockade on Qatar. The boycott decision was on 05 June 2017, which we subsequently view as the announcement day. So, this study compares the performances of these stock markets over equal periods before the boycott (Period 1: from 03 April 2016 to 04 June 2017; 428 observations), and after the blockade (Period 2: from 06 June 2017 to 07August 2018; 428 observations). We transformed all the variables by taking natural logarithms to correct for heteroskedasticity and dimensional differences. Descriptive statistics for series are reported in Table 1. Yet, at this stage (i.e., preliminary analysis), quite interesting results were drawn. We note that the volatility increased for all the stock markets under study by moving from period 1 (i.e., before the blockade, Panel A, Table 1) to period 2 (i.e., after the blockade, Panel B, Table 1), though with varying extent. The most volatile stock markets are those of Saudi Arabia and Qatar. The least volatile stock market is that of Bahrain. After the 2017 Gulf crisis, we notice that all the equities are likely to be negatively skewed, with the exception of Bahrain. Such heterogeneity in this times of market stress highlight that market participants may enjoy portfolio diversification opportunities.

**Table 1.** Statistical properties of country-level stock returns:
Before and after the blockade on Qatar

|  | QATAR | SAUDI ARABIA | UAE | BAHRAIN | EGYPT |
|---|---|---|---|---|---|
| *Panel A : Period 1 : Before the blockade on Qatar* | | | | | |
| Mean | 1.10E-05 | -0.001023 | 0.002128 | 0.001916 | 0.000190 |
| Median | -0.002028 | 0.077655 | 0.031826 | 0.031738 | 0.010806 |
| Maximum | 0.438927 | 1.677135 | 0.387325 | 0.166263 | 0.804977 |
| Minimum | -0.338575 | -4.582749 | -0.823530 | -0.698647 | -0.981078 |
| Std. Dev. | 0.181631 | 0.374123 | 0.162142 | 0.123620 | 0.226584 |
| Skewness | 0.244617 | -5.185766 | -1.539648 | -1.933086 | -0.400171 |
| Kurtosis | 4.225992 | 58.77320 | 6.920278 | 8.157172 | 5.448237 |
| Jarque-Bera | 31.07290 | 57391.57 | 443.1699 | 740.8627 | 118.3137 |
| Probability | 0.000000 | 0.000000 | 0.000000 | 0.000000 | 0.000000 |
| *Panel B : Period2 : Afterthe blockade on Qatar* | | | | | |
| Mean | 0.000316 | 0.003419 | -0.000778 | 0.000652 | 0.000171 |
| Median | -0.003971 | 0.072908 | 0.042192 | 0.039204 | 0.008809 |
| Maximum | 0.537433 | 1.684439 | 0.535590 | 0.164509 | 0.853528 |
| Minimum | -0.534400 | -4.278205 | -1.631654 | -0.456515 | -0.942518 |
| Std. Dev. | 0.297125 | 0.337480 | 0.214524 | 0.125851 | 0.239081 |
| Skewness | - 0.603860 | -5.357073 | -2.740098 | 1.122593 | -0.270353 |
| Kurtosis | 4.970826 | 65.29266 | 16.14385 | 3.759154 | 5.054545 |
| Jarque-Bera | 70.03692 | 71247.18 | 3616.481 | 100.1730 | 80.49105 |
| Probability | 0.000000 | 0.000000 | 0.000000 | 0.000000 | 0.000000 |

Fig 1 confirms that the stock price indices for most countries (especially, Qatar, Saudi Arabia and the UAE) become more volatile after the blockade in Qatar.

**Fig 1.** Stock market returns by country: Before and after the blockade

*Panel A. Period 1: Before the blockade on Qatar*

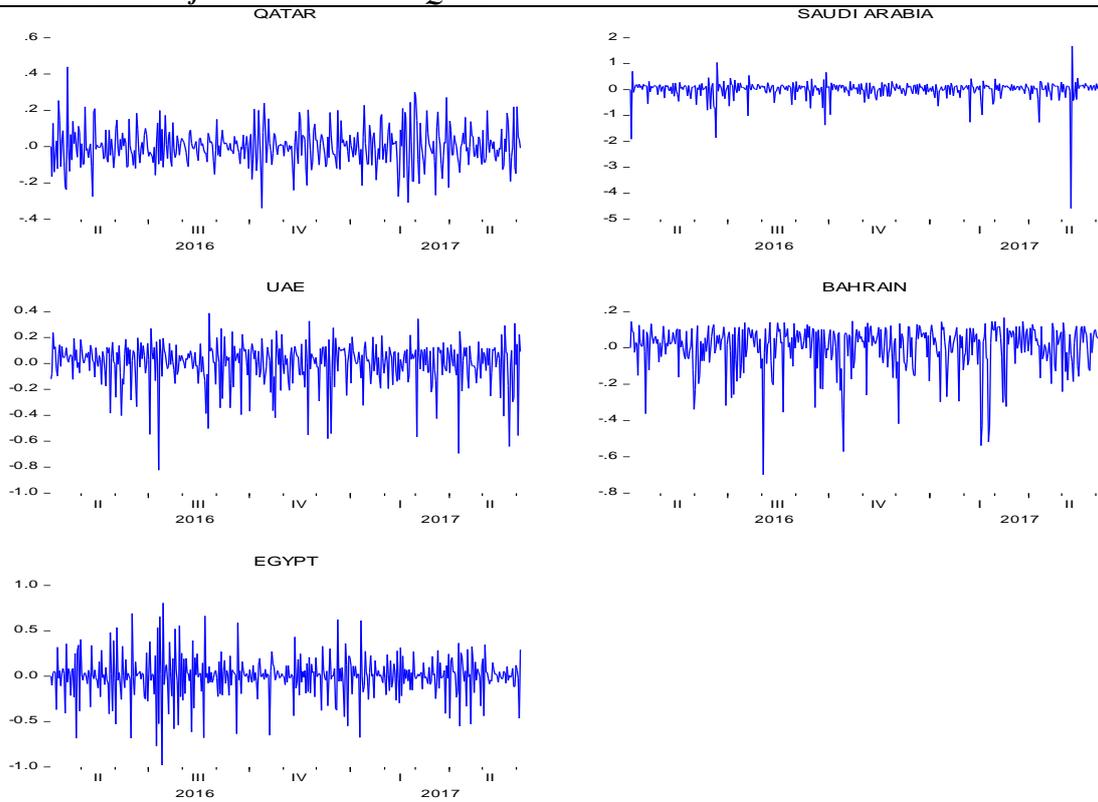

*Panel B. Period 2: After the blockade on Qatar*

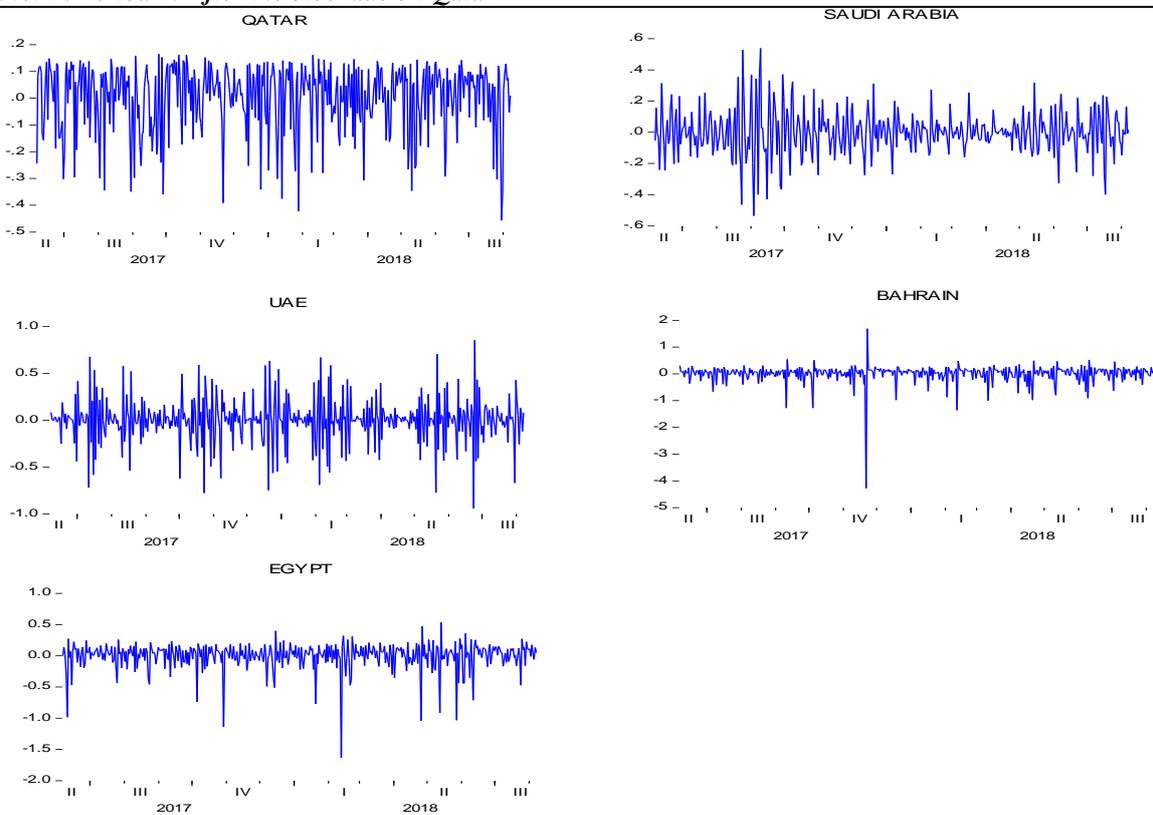

## 4. Empirical results

### *4.1. Volatility*

To choose the best GARCH model able to measure the volatilities of Qatar, Saudi Arabia, UAE, Bahrain and Egypt's stock indices, we use the Akaike information criterion. Based on this criterion, the optimal GARCH extensions chosen to capture the volatility of Qatar stock price index is the standard GARCH model for the period 1 and the Exponential GARCH model for the period 2.[5] The GARCH-type modeling has been and continues to be very valuable tool in finance and economics since the seminal paper of Engle (1982). Engle (1982) proposed to model time-varying conditional variance with Auto- Regressive Conditional Heteroskedasticity (ARCH) processes using lagged disturbances. He argued that a high ARCH order is required to properly capture the dynamic behavior of conditional variance. The Generalized ARCH (GARCH) model of Engle and Bollerslev (1986) fulfills this requirement as it is based on an infinite ARCH specification which minimizes the number of estimated parameters, denoted as:

$$\sigma_t^2 = \omega + \sum_{i=1}^{q} \alpha_i \varepsilon_{t-i}^2 + \sum_{i=1}^{p} \beta_j \sigma_{t-j}^2 \tag{5}$$

where $\alpha_i$, $\beta_i$ and $\omega$ are the parameters to estimate.

The Exponential-GARCH model introduced by Nelson (1991) contributes to the standard GARCH model by allowing to control for asymmetry. This model specified the conditional variance in a logarithmic form:

$$\log(\sigma_t^2) = \omega + \sum_{i=1}^{q}(\alpha_i z_{t-i} + \gamma_i (|z_{t-i}| - \sqrt{2/\pi})) + \sum_{i=1}^{p} \beta_j \log(\sigma_{t-j}^2) \tag{6}$$

---

[5]The detailed Akaike information criterion results will be available for interested readers upon request.

where $\alpha_i$, $\beta_j$, $\omega$, $\gamma$ are the parameters to estimate, and $z_t$ the standardized value of error.

For Saudi Arabia, the optimal model based on the AIC information criterion able to capture best the stock market index volatility is the Threshold-GARCH model for the two periods (before and after the 2017 Gulf crisis). The Threshold-GARCH developed by Zakoin (1994) accommodates structural breaks in volatility. It allows describing the regime shifts in the volatility, denoted as:

$$\sigma_t^2 = \omega + \sum_{i=1}^{q}(\alpha_i|\varepsilon_{t-i}| + \gamma_i|\varepsilon_{t\_i}^+|) + \sum_{i=1}^{p}\beta_j\sigma_{t-j} \qquad (7)$$

where $\alpha_i$, $\beta_j$, $\omega$ and $\gamma$ are the parameters to estimate.

For the UAE and Egyptian stock indices, the most appropriate GARCH model selected based on the same information criterion is the Exponential-GARCH model for the period 1 and the Threshold-GARCH model for the period 2.

For Bahrain stock price index, the Integrated-GARCH model seems the most appropriate volatility measure for period 1, while the Threshold-GARCH is the best volatility indicator for period 2. In many analyses of the variables behaviour of volatility, a vexing question regards the persistence of long shocks to conditional variance. The Integrated GARCH model is a part of a large class of models with a property called "persistent variance", which assumes that current information is still substantial for the forecasts of the conditional variances for all time horizons.

$$\sigma_t^2 = \omega + \varepsilon_{t-1}^2 + \sum_{i=1}^{q}\alpha_i(\varepsilon_{t-i}^2 - \varepsilon_{t-1}^2) + \sum_{i=1}^{p}\beta_j(\sigma_{t-j}^2 - \varepsilon_{t-1}^2) \qquad (8)$$

where $\alpha_i$, $\beta_j$, $\omega$ and $\gamma$ are the parameters to estimate.

The estimates are reported in Table 2. Our results indicate that the volatile behaviors of the stock price indices for all the countries under study change slightly by moving from the period prior to the Qatar crisis (period 1; Panel A, Table 2) to the post-boycott (period 2; Panel B, Table 2). All the stock markets become more volatile in response to the blockade, but such volatility does not persist. In particular, the duration of persistence is far from one for all cases, and thus we did not find any evidence of long memory in the conditional variance. The asymmetrical effect is positive and statistically significant for all the considered stock markets implying that the effect of bad news on the conditional variance exceeds that of good news. Indeed, the degree of asymmetry ($\frac{\alpha+\gamma}{\alpha}$), which measures the relative influence of bad news on volatility seems important for the majority of cases (it amounts 1.00 for all cases). The degree of asymmetry is still pronounced for the two periods, confirming the moderate effect of Qatar diplomatic crisis on Gulf region equity markets.

**Table 2.** Volatility' parameters by country: Before and after the blockade on Qatar

|  | QATAR | SAUDI ARABIA | UAE | BAHRAIN | EGYPT |
|---|---|---|---|---|---|
| *Panel A: Period 1:Before blockade on Qatar* | | | | | |
| | | Mean equation | | | |
| $C$ | -0.013*** | 0.0272 | 1.8134*** | -0.328* | -0.413*** |
| | (0.0007) | (0.2464) | (0.0000) | (0.0567) | (0.0001) |
| Lagged returns | 0.1752*** | -0.0723 | 0.127*** | 0.155*** | 0.139* |
| | (0.0000) | (0.4299) | (0.0002) | (0.0000) | (0.040) |
| | | Variance equation | | | |
| $\omega$ | 0.0007*** | 0.272*** | 0.214* | 0.311* | 0.204*** |
| | (0.0004) | (0.0000) | (0.0362) | (0.0104) | (0.0009) |
| $\alpha$ | -0.042*** | 0.728*** | 0.441 | 0.076** | 0.023** |
| | (0.0000) | (0.0000) | (0.8229) | (0.0055) | (0.0055) |
| $\beta$ | 0.6354*** | -0.008 | 0.221* | 0.571** | 0.514** |
| | (0.0000) | (0.8445) | (0.0303) | (0.0034) | (0.0026) |
| $\gamma$ | --- | 0.001* | 0.016*** | --- | 0.0002* |
| | | (0.0114) | (0.0000) | | (0.0153) |
| The duration of persistence: $\alpha+\beta+0.5\gamma$ | 0.59 | 0.72 | 0.74 | 0.64 | 0.49 |
| The leverage effect: $\gamma$ | --- | 0.001 | 0.016 | --- | 0.0002 |
| *Panel B: Period 2:After blockade on Qatar* | | | | | |
| | | Mean equation | | | |
| $C$ | 0.0912*** | 0.401*** | 0.748* | 0.338 | 0.293** |
| | (0.0003) | (0.0000) | (0.0617) | (0.3371) | (0.0014) |
| Lagged returns | -0.0634* | -0.1032* | 0.354*** | -0.4214* | -0.4256** |
| | (0.0271) | (0.0218) | (0.0003) | (0.0124) | (0.0078) |

|  | Variance equation | | | | |
| --- | --- | --- | --- | --- | --- |
| $\omega$ | 0.0145** | 0.0166* | -0.632*** | 0.0451* | 0.0452* |
|  | (0.0059) | (0.0414) | (0.0000) | (0.0310) | (0.0357) |
| $\alpha$ | 0.368*** | 0.3019** | 0.7839*** | 0.130** | 0.030** |
|  | (0.0005) | (0.0038) | (0.0000) | (0.0036) | (0.0036) |
| $\beta$ | 0.352** | 0.5107 | 0.0145*** | 0.533*** | 0.418*** |
|  | (0.0044) | (0.1349) | (0.0000) | (0.0004) | (0.0004) |
| $\gamma$ | 0.0007*** | 0.0012* | 0.0004*** | 0.001** | 0.031 |
|  | (0.0000) | (0.0103) | (0.0000) | (0.0672) | (0.211) |
| The duration of persistence: $\alpha + \beta + 0{,}5\gamma$ | 0.73 | 0.81 | 0.79 | 0.67 | 0.51 |
| The leverage effect: $\gamma$ | 0.0007 | 0.0012 | 0.0004 | 0.001 | 0.031 |

Notes: $\omega$: the reaction of conditional variance; $\alpha$: the ARCH effect; $\beta$: the GARCH effect; $\gamma$: the leverage effect;(.): the p-value; p-value<0.01: ***; p-value<0.05: **; p-value<0.1:*.With respect to the results of AIC information criterion, we select one lag for all the specifications.

The conditional variances processes displayed in Fig 2 indicate that the persistence of stock market volatility differs substantially from one country to another and from the period before the boycott to the period post-blockade. After the boycott, the conditional variance appears more persistent in Qatar, Saudi Arabia and the UAE.

**Fig. 2.** Conditional variance of stock returns by country: Before and after the blockade on Qatar

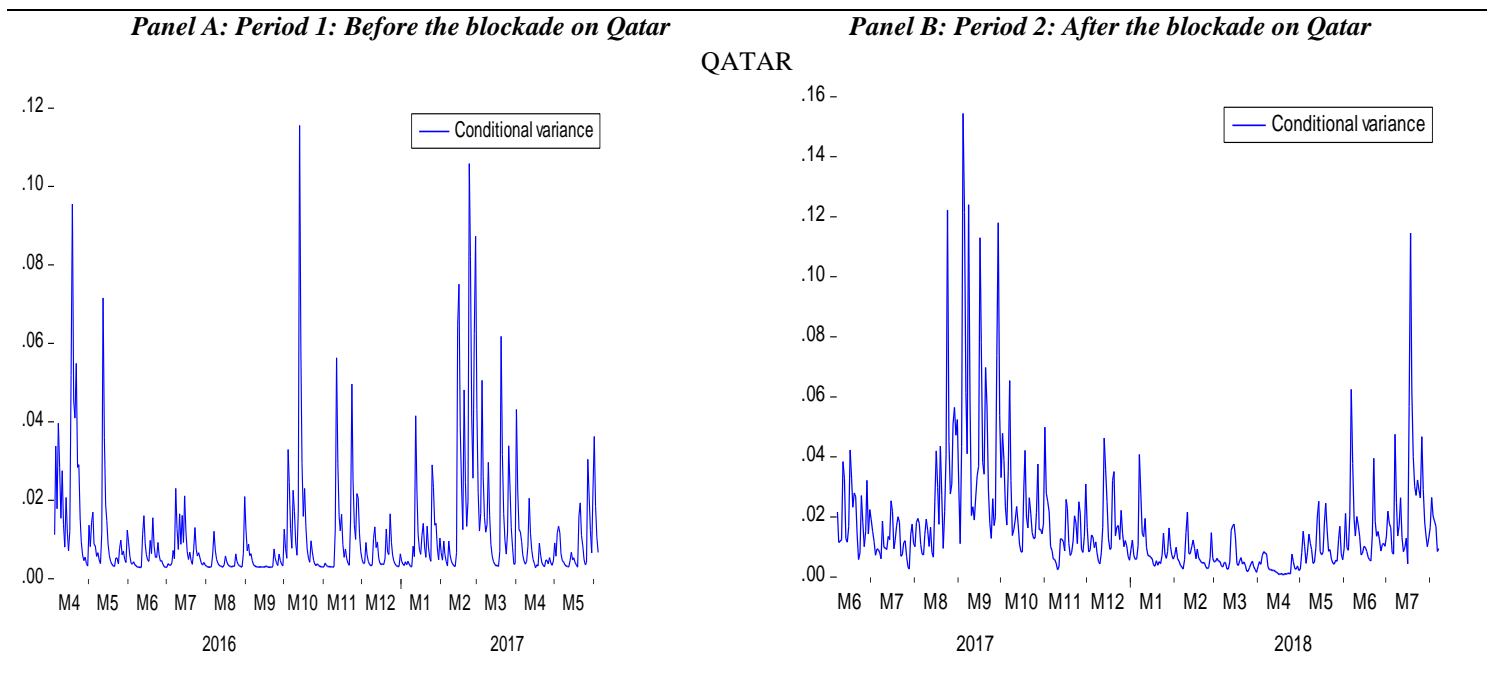

SAUDI ARABIA

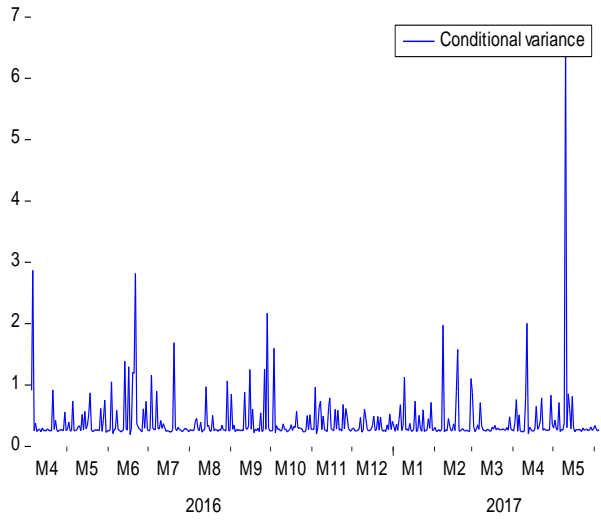
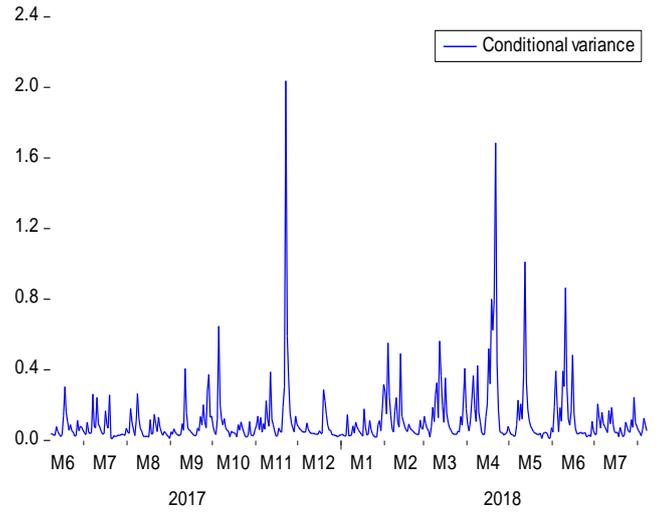

UAE

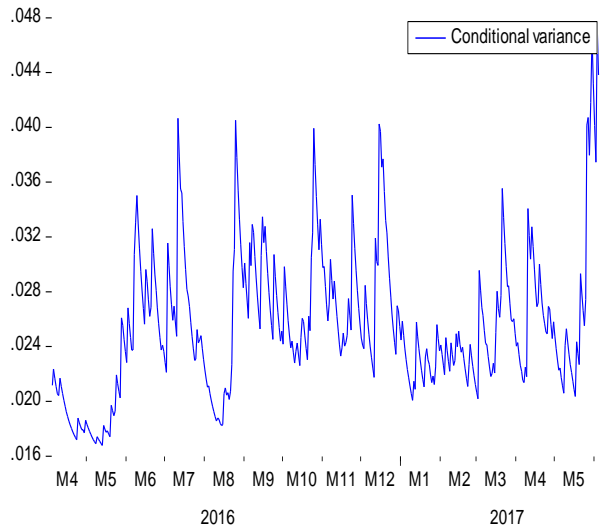
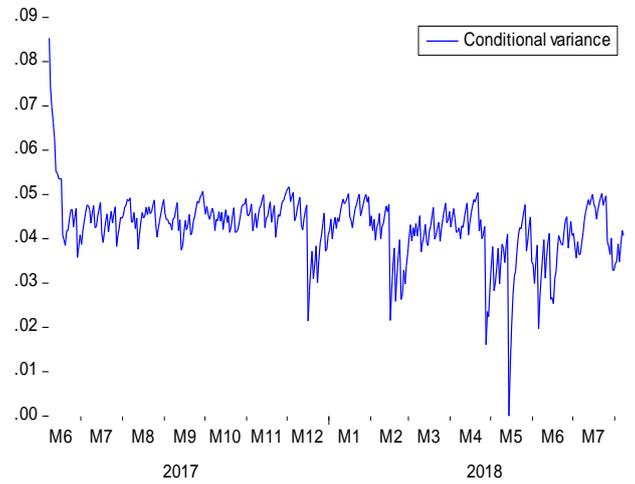

BAHRAIN

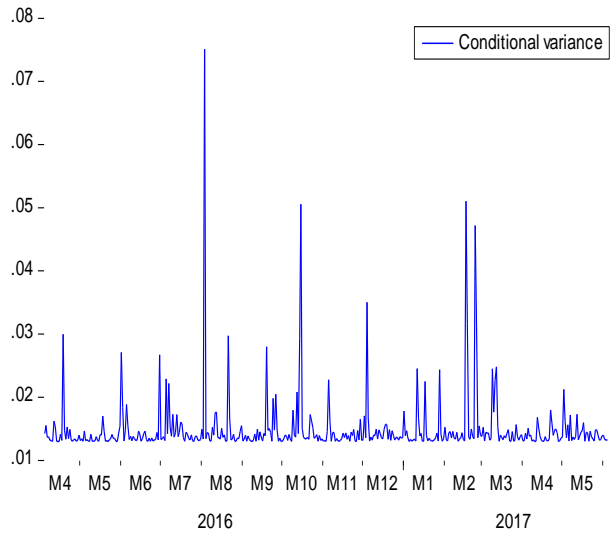
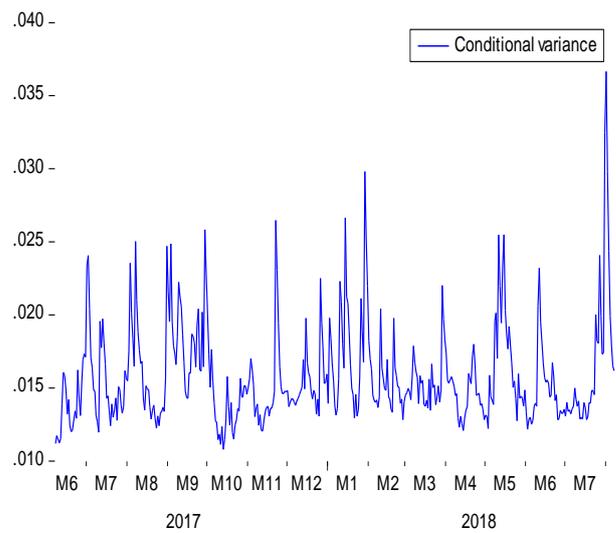

EGYPT

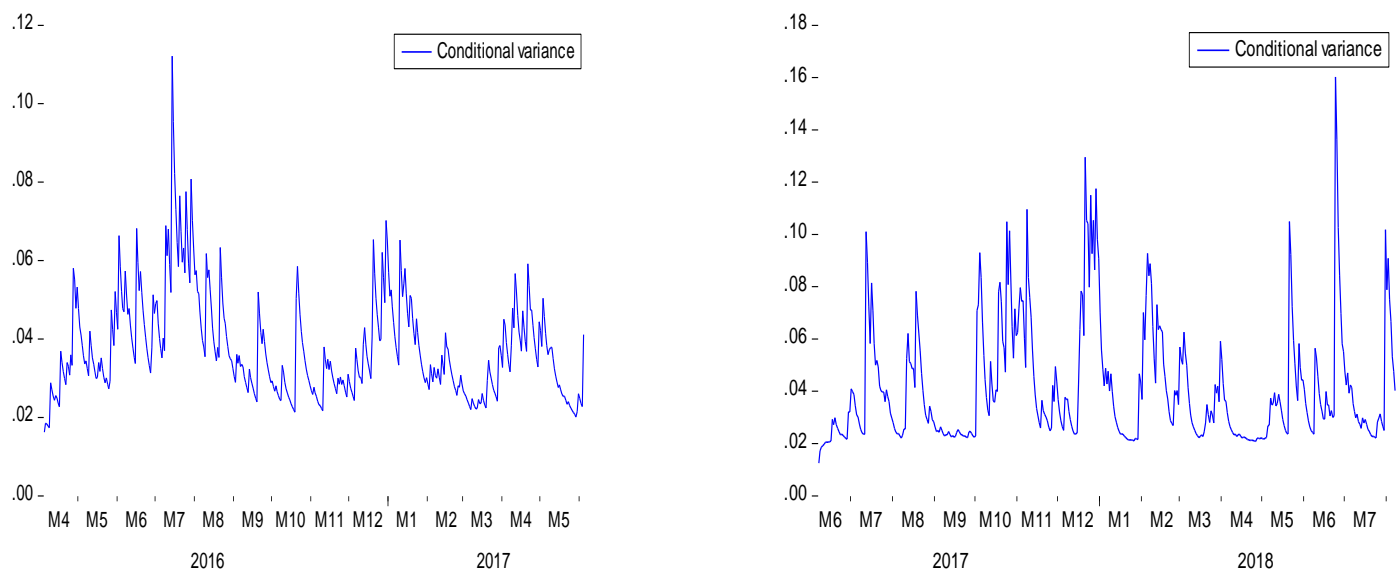

For comparison purpose, we tested the effect of this crisis on Kuwait and Oman stock markets. This would allow us to assess whether a neutral reaction may help to avoid volatility spillovers. Kuwait has attempted to mediate the spat between Qatar and its Gulf neighbors. Its good links with all parties of the GCC and equal distance from each of them have enabled Kuwait to act in a neutral manner. Oman is uninvolved in the 2017 Gulf crisis and cannot undertake such a mission because of tense relations with Saudi Arabia and the UAE as a consequence of strong Oman's ties with Iran. From Table A.2 and Fig A.2 (preliminary results), we note that the Kuwaiti and Oman's stock markets do not change fundamentaly by moving from the period prior to the blockade to the post-boycott period. The volatility increase modestly after the blockade on Qatar. We select then the best optimal model for each stock price index based on AIC information criterion. The findings derived from the optimal GARCH model of each stock market (Table A.3, Appendix) reveal that the crisis affect modestly the volatility of stock markets. We note a relatively moderate increase in the duration of persistence. Fig A.3 (Appendix) confirm that the volatility increase weakly after the blockade. For the two periods, the Kuwaiti stock market and Muscat shares seem more responsive to good news (i.e., negative leverage effect ; see Table A.3).

## 4.2. Volatility spillovers across Qatar and the boycotting countries

In the aftermath of a sudden political decision, such as the boycott against Qatar, the associated ramifications on the stock markets, particularly the regional ones, are questionable. In addition to the investigation of the effect of the 2017 Qatar diplomatic crisis on the volatility, speculative attitude and the efficiency of Qatar, GCC and Egyptian stock markets, we assess the financial spillover effect of the regional turmoil on Qatar and the boycotting countries. Table 4 summarizes an approximate "input-output" decomposition of the total volatility spillover index. In particular, based on the study of Diebold and Yilmaz (2012), we decompose the spillover index into all of the forecast error variance components for variable $i$ coming from shocks to variable $j$, for all $i$ and $j$. The $ij^{th}$ entry is the estimated contribution to the forecast variance of market $i$, resulting from innovations to market $j$. The sum of variances in a row (column), excluding the contribution to its own volatilities (diagonal variances) corresponds to the effect on the volatilities of other stock markets. The last row in the table is the contribution to the volatilities of all markets from this particular market.

Before the 2017 Qatar-Gulf crisis (Panel A, Table 3), the volatility spillovers to others (98.3%) is greater than the volatility spillovers from others (59.2%). After the blockade on Qatar (Panel B, Table 3), we clearly note that the volatility transmission to and from others increase but not strongly. In particular, our results reveal that for total volatility spillovers to others (107.7%) is stronger than total volatility spillovers from others (63.4%). For Qatar, Saudi Arabia and the UAE, the contribution to others is more important than the contribution from others; inversely for Bahrain and Egypt. This holds true for the two periods under study. The important volatility transmission among GCC markets before and after the blockade can be explained by the increased financial sector integration among Gulf countries. Highly motivated by the necessity to enhance efficiency, GCC countries have taken prominent steps these last decades toward achieveing appropriate financial regulation and corporate governance measures, which have in turn enabled to improve convergence across GCC financial systems. Lane and Milesi-Ferretti (2017) explored the extent of financial integration in the Gulf using capital flow data and equity prices. The study revealed that there is some improvement in regional financial integration. Although the Qatar diplomatic crisis has intensified the volatility spillovers, this effect does not appear pronounced. Even modestly, we note an increased risk spillover among Qatar, GCC and Egyptian stock markets by moving from period 1 (before the blockade) to period 2 (after the blockade). This can be viewed as a signal of limitations of portfolio diversification opportunities during this crisis period.

**Table 3.** Stock market volatility spillovers across Qatar and the boycotting countries: Before and after the blockade on Qatar

|  | Qatar | Bahrain | Saudi Arabia | UAE | Egypt | Contribution from others |
|---|---|---|---|---|---|---|
| *Panel A. Period 1: Before the blockade on Qatar* | | | | | | |
| Qatar | 58.7 | 7.3 | 14.5 | 12.7 | 3.6 | 8.6 |
| Bahrain | 8.9 | 31.4 | 9.2 | 5.9 | 4.9 | 14.3 |
| Saudi Arabia | 31.4 | 4.6 | 51.4 | 17.1 | 2.5 | 6.5 |
| UAE | 7.4 | 5.1 | 8.1 | 41.7 | 1.6 | 5.9 |
| Egypt | 1.9 | 3.4 | 6.7 | 8.4 | 40.3 | 12.9 |
| Contribution to others | 19.8 | 9.8 | 26.0 | 24.2 | 7.2 | 59.2 |
| Contribution including own | 79.8 | 41.2 | 77.4 | 65.9 | 47.5 | 36.8 |
| *Panel B. Period 2: After the blockade on Qatar* | | | | | | |
| Qatar | 63.9 | 9.7 | 23.4 | 14.9 | 4.2 | 11.9 |
| Bahrain | 5.1 | 36.5 | 8.7 | 6.6 | 3.4 | 19.3 |
| Saudi Arabia | 10.3 | 1.3 | 62.1 | 12.3 | 1.9 | 8.1 |
| UAE | 9.7 | 4.5 | 7.3 | 55.9 | 1.3 | 6.8 |
| Egypt | 2.7 | 2.0 | 11.9 | 9.3 | 61.5 | 17.3 |
| Contribution to others | 53.4 | 7.2 | 20.9 | 17.6 | 8.6 | 63.4 |
| Contribution including own | 117.3 | 102.7 | 83.0 | 73.5 | 70.1 | 43.6 |

Notes: The values are calculated from variance decompositions based on 1-step-ahead forecasts. The optimal lag length for the VAR models is 3 for the two periods under study, determined by the Akaike Information Criterion.

Thereafter, we determine the average net directional spillovers prior to and post-the Qatar diplomatic crisis, which is the difference between the "contribution to others" and the "contribution from others". This task permits to identify which from the stock markets under study is the most potential in exporting volatilities to the other countries during the boycott against Qatar. The results are reported in Table 4. We show that the results change but not fundamentally after the recent Gulf crisis. Before the boycott, two groups of countries are derived: Qatar, Saudi Arabia and the UAE are viewed as volatility transmitters; while Bahrain and Egypt are considered as risk receivers (Panel A, Table 4). After the crisis, we keep the same groups of countries, though with changing intensity of volatility spillovers. In particular, with an average net directional return spillover of 41.5%, the Qatar stock market appears the most influential in transmitting risk to others countries (Panel B, Table 4), followed by Saudi Arabia (12.8%) and UAE (10.8%). Nevertheless, the stock markets of Bahrain and Egypt - with negative volatility spillover indexes (-12.1% and -8.7%, respectively) - are regarded as net volatility receivers. The identification of volatility transmitters and receivers may help in designing effective hedging strategies. Investors can enhance their hedging and portfolio diversification by exploiting its knowledge with respect the way the risks associated to stock markets over the Qatar diplomatic crisis can be transmitted from one market to another. As hopes of swift resolution to the standoff seem increasingly remote, providing useful information regarding the directional spillovers should allow regulators undertake preventive strategies to mitigate the volatility transmission from the Qatar, Saudi Arabia and UAE to Bahrain and with less extent Egypt. This requires an effective management of financial risks by ensuring adequate regulation and supervision (Caffagi and Miller 2013).

**Table 4.** The average net directional volatility spillovers across Qatar and the boycotting countries: Before and after the blockade on Qatar

|  | Contribution from others | Contribution to others | Average net directional spillover |
|---|---|---|---|
| *Panel A. Period 1: Before the blockade on Qatar* | | | |
| Qatar | 8.6 | 19.8 | 11.2 |
| Bahrain | 14.3 | 9.8 | -4.5 |
| Saudi Arabia | 6.5 | 26.0 | 20.5 |
| UAE | 5.9 | 24.2 | 18.3 |
| Egypt | 12.9 | 7.2 | -5.7 |
| *Panel B. Period 2: After the blockade on Qatar* | | | |
| Qatar | 11.9 | 53.4 | 41.5 |
| Bahrain | 19.3 | 7.2 | -12.1 |
| Saudi Arabia | 8.1 | 20.9 | 12.8 |
| UAE | 6.8 | 17.6 | 10.8 |
| Egypt | 17.3 | 8.6 | -8.7 |

To ascertain the robustness of these results, we incorporated in the vector autoregressive (VAR) model, the equities of Kuwait and Oman. In doing so, the results remain robust to total volatility spillovers to others are still more pronounced than risk spillovers from others. We also confirm that the effect of Qatar crisis on the volatility transmission is relatively low (see Table A.4, Appendix). In addition, Qatar, Saudi Arabia and the UAE remain net volatility transmitters, while Bahrain, Egypt, Kuwait and Oman are considered as volatility receivers (see Table A.5, Appendix).

### *4.3. Discussion of results: Heightened diplomatic tensions with limited economic repercussions*

The analysis carried out showed that there is a real competition between the different countries for the regional leadership, and they each have strengths and limitations. Saudi Arabia can appear as a giant compared to other Gulf countries. However, this asymmetry is only apparent. Much diplomatic maneuvering succeeded in bringing a small state to convert a crisis targeting its leadership and sovereignty and aiming to eliminate its independence, and to successfully deal with economic uncertainty. This unprecedented crisis will escalate tensions between the protagonists in the region that is, by nature, very unstable. Our results reveal that while Qatar has been shaken by this crisis, the other countries are not left out, especially Saudi Arabia and the UAE. We try in the following to provide some answers to these questions: What are the main Qatar's elements of strength? What are the regional and global factors of resilience that helped Qatar resist the blockade? How the Saudi-led blockade failed to achieve its goals?

- *(i)* *Qatar's economic resilience:* The Qatar diplomatic dispute is the biggest political crisis to hit the Middle East in several years. The quartet has tried to strangle the Qatar's economy through an unprecedented blockade in the recent history. More than one year ago, an air, sea and land blockade was imposed on Qatar. Certainly, this blockade is not without consequences for this small country, especially that it was unprepared for such a major escalation. Qataris are likely to find it very difficult (if not impossible) to import their basic needs. Qatar is hugely dependent on imports by land and sea, and approximately 40 percent of its food came in via the land border with Saudi Arabia. At the same time, the crisis has put pressure on the Qatari riyal, and the country has been enforced to dip into its reserves to preserve its currency's value against the dollar.[6] Car sales also witnessed a gradual downward trend after the announcement of boycott. Likewise, the crisis has increasingly affected the tourism industry. Further, Qatar's efforts to fight the ongoing blockade have worsened the budget deficit. In brief, the flashing lights of economic indicators were all red. However, Qatar's wider reliance on extractive hydrocarbon resources allowed the country to conduct an active foreign policy. After a period of rising uncertainty, the Qatari authorities responded vigorously and quickly to the blockade. Since blockade imposed against Qatar, new maritime and air trade routes have opened, especially to Iran, Turkey and Pakistan. On the local scene, before the boycott, the local production covered 15 percent of domestic demand Qatar imported 80 percent of its food needs from Saudi Arabia and the UAE. After the blockade, Qatar plans to limit food imports by 60 per cent by adopting innovative production technologies to grow agricultural products in an effort to meet the market demand. Against the backdrop of the blockade, Qatar has proved that business is open as ever. In fact, the Qatari airline, surrounded by everywhere, has experienced rapid expansion this year, proudly standing as one of the world's fastest-growing airlines. It has also announced the opening of twenty new destinations. Similarly, see ports of Qatar, largely under-exploited until this blockade, have witnessed an increased growth. Also, the Qatar's energy sector has shown greater resilience, adaptability and determination to lessen the harmful impact of the crisis. According to the 2018 world Liquefied Natural Gas (LNG) report released by International Gas Union, Qatar has retained its position as world's major LNG exporter in 2017 in a sign of strength amid Gulf rift. Recently, Qatar has announced a biggest deal to supply liquefied natural gas to Beijing for the next twenty years. China will buy 3.4 million tonnes every year from Qatar. This would help to largely improve the Qatar's position as a leading natural gas exporter over the long-run. Add to this that Qatar's economic growth is expected to recover to 2.8 per cent in 2018, and increase further to attain an average of 3 per cent in 2019, as growing energy receipts allow easing fiscal constraints (World Bank report, 2018). Qatar also adopted a new in 2017 which offers legal guarantees for domestic workers' labor rights. This significant change is part of Doha's efforts to enhance international perception of this small state as it seeks to fight the diplomatic isolation and escalated pressures from its neighbors. If this blockade showed the resilience of Qatar's economy, it also highlights the incapacity of the boycotting countries to put it down.

---

[6] Qatar has $340 billion in reserves that could help the Gulf country to circumvent the isolation by its neighboring countries.

(ii) *Saudi Arabia- A giant with feet of clay:* The Saudi economy is the largest in the Arab world. It is highly dependent on oil. This country has the world's second-largest proven petroleum reserves after Venezuela and it is the largest exporter of petroleum. Add to this, Saudi Arabia has the fifth-biggest proven natural gas reserves. Saudi Arabia is commonly regarded as an energy superpower. But since the 2014 oil price decline, the country is plagued by major economic hardships, which has forced it to reduce its public spending. Oil is still account for about 80 per cent of Saudi exports, and three-quarters of total tax revenue depend on it. The serious oil price collapse forced Saudi Arabia to undertake deeper changes to its economy. The Saudi government has imposed new taxes, including a 5 percent value added tax (VAT). It must be stressed that this is the first tax imposed in the country. The country has also accelerated its efforts to build a more diversified industrial economy, with new facilities for various sectors including chemicals, fertilizers, aluminum and cement. Regardless of Saudi Crown Prince's unprecedented reform efforts, shifting to a diversified economic structure seems not easier for Saudi Arabia. This is attributed, even partially, to the fact that Saudi Arabia, as a "rentier state" and therefore, has had a limited incentive to spur the growth of any non-oil sector of its economy. Another major shift in Saudi Arabia could be the partial privatization of Saudi Aramco. Based on Bloomberg news, Saudi Aramco is one of the top-companies in the world by revenue. It is the world's second-largest proven oil reserves, at more than 270 billion barrels. Accordingly, International Monetary Fund proclaims Saudi economy, which contracted by 0.9 per cent in 2017, is expected to grow by 2.2 per cent in 2018 and 2.4 per cent in 2019. However, the rise in the price of black gold will be insufficient to relieve the social pressures in Saudi Arabia, partly fueled by an increase in unemployment among young people under the age of 20 to 24 (42 per cent).Companies operating or planning to invest in Saudi Arabia face also a significant risk of corruption.[7] The privatization of Saudi Aramco, which constitutes the barley point of this strategy of seduction, indefinitely postponed, according to Saudi sources. It is also difficult to attract foreign investors when Saudi officials do not provide information about the volume of reserves of proven oil reserves. Likewise, the company's accounts have never been audited. For boosting international investors' confidence and for Saudi Arabia's economic reforms to carry credibility, there is an urgent necessity for greater transparency in how government finances are generated and dispersed. According to the UN Conference on Trade and Development (UNCTAD, 2017), inward investment into Riyadh dropped markedly in 2017, raising several questions regarding the prospects for the economic reform agenda being conducted by Crown Prince MBS. The case of the Saudi journalist killed inside the Saudi consulate in Istanbul, harmed kingdom image and will keep foreign investors' fingers mightily on the pause button when it comes to allocating to the country.[8]

---

[7]Corruption has long been endemic in Saudi Arabia, According to the 2017 Corruption Perceptions Index, corruption rank in Saudi Arabia averaged 62.47 from 2003 until 2017.

## 5. Conclusions

There is bountiful evidence that political uncertainty make financial markets significantly volatile. Accordingly, Poon and Granger (2003) argued that precise prediction of volatility is highly prominent for at least four reasons. First, when the volatility is interpreted as uncertainty, it becomes a potential input to make appropriate investment decisions and portfolio allocation. Second, analyzing the volatility dynamics is of paramount importance in the pricing of derivative securities. Third, financial risk management necessitates an effective prediction of volatility as a requisite input to risk management for financial institutions (Rapach et al. 2008 ; Gil-Alana et al. 2014 ; Yaya et al. 2015). Fourth, the equity market volatility can have large repercussions on the economy as a whole through its impact on real economic activity and public confidence. Certainly, estimates of market volatility during periods of rising uncertainty can be perceived as efficacious measure for the vulnerability of financial markets and the economy, and can allow policymakers designing the best possible policies. In short, a good prediction of the process of volatility has relevant implications for investment decisions, portfolio allocation, the pricing of derivative securities and risk management.

---

[8] The Saudi equity market dropped by about 7.2 per cent between October 10-14 while the story was developing.

Given this, the present study seeks to examine the impact of the coalition of Arab countries led by Saudi Arabia imposed a historic land, maritime, and air blockade on the stock market volatility in the Gulf region and risk spillovers across these markets. Despite our awareness that it is difficult to quantify with certainty the costs of 2017 Gulf crisis, our estimations give quite interesting insights. The economic implications of the Qatar's isolation are likely to be costly but short-lived. The GCC crisis has inflicted significant financial loss not only on Qatar but also on the boycotting countries (i.e., a lose-lose scenario). Specifically, our findings indicate that the equities of Qatar, Saudi Arabia, UAE, Bahrain and Egypt become more volatile and relatively more responsive to bad news. But this volatility does not persist. Our findings also document that the profound political instability over Qatar crisis weakly exacerbate the stock market volatility transmission across Qatar and the boycotting countries. In short, our results suggest that the boycott did not achieve the expected outcome. The fact that Qatar, Saudi Arabia and the UAE responded in the same way (with respect the volatility persistence and the directional risk spillovers) to this crisis can be considered as a sign that Qatar "beats" the boycott. Doha has demonstrated resilience in times of heightened political uncertainty. Despite its economic vulnerability, Qatar has successfully resisted the Saudi-embargo. More than sixteen months later, the blockaders show no signs of relenting. Many factors can explain Qatar's model of resisting blockade. Using income from its wide gas reserves to bankroll its ambitious plans, Qatar has carved out a powerful regional and global profile in the past decade, and has been perceived as significant power in the Arab world. In response to the blockade, Qatar rebuild its trade ties and food supply chain to pass round imports previously received through Saudi Arabia and the UAE. Qatar has also retained the crown of world's top exporter of liquefied natural gas in 2017, underpinning Qatari cash flow. Further, Qatar withstand the harmful effects of the blockaders it growingly emphasizes economic relationships outside the Gulf region. This has allowed Doha to replace import trade that came by land from Saudi Arabia and by sea from the UAE. Overall, the resilience of this tiny state appears as a model on how turning crisis into opportunity. Even though Doha has a long-term plan to become less dependent upon gas revenues, there was still a strong reliance on supply routes and potential trade partners. The 2017 Gulf crisis forced Qatar to think and act more swiftly to determine new supply routes and trade partners. The recent Gulf crisis and its resulted diplomatic and economic challenges with other GCC countries has significantly sped up Qatari plans and has also strengthened the motivation to take a close attention to self-sufficiency.

Last but not least, our empirical findings reveal that Qatar diplomatic crisis creates new Gulf with no winners. This crisis has further divided the Arab and Muslim world, and forced small states to make tough choices. We do not know with certitude how this diplomatic crisis will reach a climax and precisely what the long-run ramifications will be. But past imposition of boycott gives a practical exhibition of a variety of unanticipated consequences ranging from undermining the embargoing countries' diplomatic influence, to heightened political instability, to significant escalation as one or both sides would attempt to erupt a strategic stalemate (Doughty and Raugh 1991; Robbins 2013; Colins 2018). With the continuing standoff between Qatar and Saudi Arabia, diplomatic and political relationships between several Arab countries will likely suffer further damage. It is time for these countries to resolve their differences and work on strengthening the GCC macroeconomic outcomes in an uncertain global economy, which goes hand in hand with the promotion of democratization. They have the potential, but they lack the will to act.

# Appendix

**Table A.1.** GARCH models used in this study

| |
|---|
| GARCH-M (GARCH in mean, Bollerslev et al., 1993) |
| $r_t = \mu_t + \varepsilon_t + \lambda \sigma_t^2$ |
| C-GARCH (Component GARCH, Ding et al. 1993) |
| $(\sigma_t^2 - \sigma^2) = \alpha(\varepsilon_{t-1}^2 - \sigma^2) + \beta(\sigma_{t-1}^2 - \sigma^2)$ |
| I-GARCH (Integrated GARCH, Bollerslev and Engle, 1993) |
| $\sigma_t^2 = \omega + \varepsilon_{t-1}^2 + \sum_{i=1}^{q} \alpha_i(\varepsilon_{t-i}^2 - \varepsilon_{t-1}^2) + \sum_{i=1}^{p} \beta_j(\sigma_{t-j}^2 - \varepsilon_{t-1}^2)$ |
| T-GARCH (Threshold GARCH, Zakoian, 1994) |
| $\sigma_t^2 = \omega + \sum_{i=1}^{q}(\alpha_i|\varepsilon_{t-i}| + \gamma_i|\varepsilon_{t-i}^+|) + \sum_{j=1}^{p} \beta_j \sigma_{t-j}$ |
| E-GARCH (Exponential GARCH, Nelson, 1991) |
| $\log(\sigma_t^2) = \omega + \sum_{i=1}^{q}(\alpha_i z_{t-i} + \gamma_i(|z_{t-i}| - \sqrt{2/\pi})) + \sum_{j=1}^{p} \beta_j \log(\sigma_{t-j}^2)$ |
| P-GARCH (Power GARCH, Higgins and Bera, 1992) |
| $\sigma_t^\varphi = \omega + \sum_{i=1}^{q} \alpha_i \varepsilon_{t-i}^\varphi + \sum_{j=1}^{p} \beta_j \sigma_{t-j}^\varphi$ |
| A-PGARCH (Asymmetric power GARCH, Ding et al., 1993) |
| $\sigma_t^\varphi = \omega + \sum_{i=1}^{q} \alpha_i(|\varepsilon_{t-i}| + \gamma_i \varepsilon_{t-i})^\varphi + \sum_{j=1}^{p} \beta_j \sigma_{t-j}^\varphi$ |
| CMT-GARCH (Component with Multiple Thresholds GARCH) |
| $\sigma_t^2 = \omega + \alpha \varepsilon_{t-1}^2 + \beta(\omega + (\alpha + \gamma I_{(\varepsilon_{t-2}<0)})\varepsilon_{t-2}^2 + \beta \sigma_{t-2}^2)$ |

Notes: $\sigma_t^2$: conditional variance, $\alpha_0$: reaction of shock, $\alpha_1$: ARCH term, $\beta_1$: GARCH term, $\varepsilon$: error term; $I_t$: denotes the information set available at time t; $z_t$: the standardized value of error term where $z_t = \varepsilon_{t-1}/\sigma_{t-1}$; $\mu$: innovation, $\gamma$: leverage effect; $\varphi$: power parameter.

**Table A.2.** Statistical properties of Kuwaiti and Muscat stock returns:
Before and after the blockade on Qatar

| | KUWAIT | OMAN |
|---|---|---|
| *Panel A : Period 1 : Before the blockade on Qatar* | | |
| Mean | 0.000806 | 0.001621 |
| Median | 0.030342 | 0.041192 |
| Maximum | 0.116793 | 0.145847 |
| Minimum | -1.158281 | -0.509544 |
| Std. Dev. | 0.108400 | 0.125771 |
| Skewness | -4.106304 | -1.801694 |
| Kurtosis | 7.85176 | 6.169026 |
| Jarque-Bera | 192.9533 | 41.06505 |
| Probability | 0.000000 | 0.000000 |
| *Panel B : Period 2 : After the blockade on Qatar* | | |
| Mean | 0.000338 | -0.001613 |
| Median | 0.040166 | 0.041048 |
| Maximum | 0.127914 | 0.133137 |
| Minimum | -1.625613 | -0.589045 |
| Std. Dev. | 0.131264 | 0.123171 |
| Skewness | -5.588247 | -1.676758 |
| Kurtosis | 5.964327 | 5.930341 |
| Jarque-Bera | 59.44518 | 35.36880 |
| Probability | 0.000000 | 0.000000 |

**Fig A. 1.** The evolution of Kuwaiti and Muscat stock market returns:
Before and after the blockade

*Panel A. Period 1: Before the blockade on Qatar*

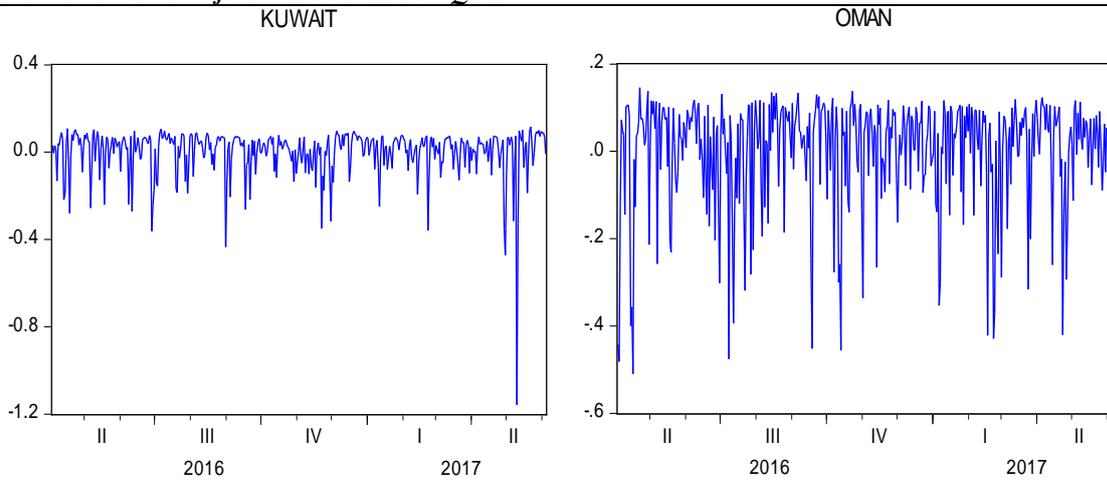

*Panel B. Period 2: After the blockade on Qatar*

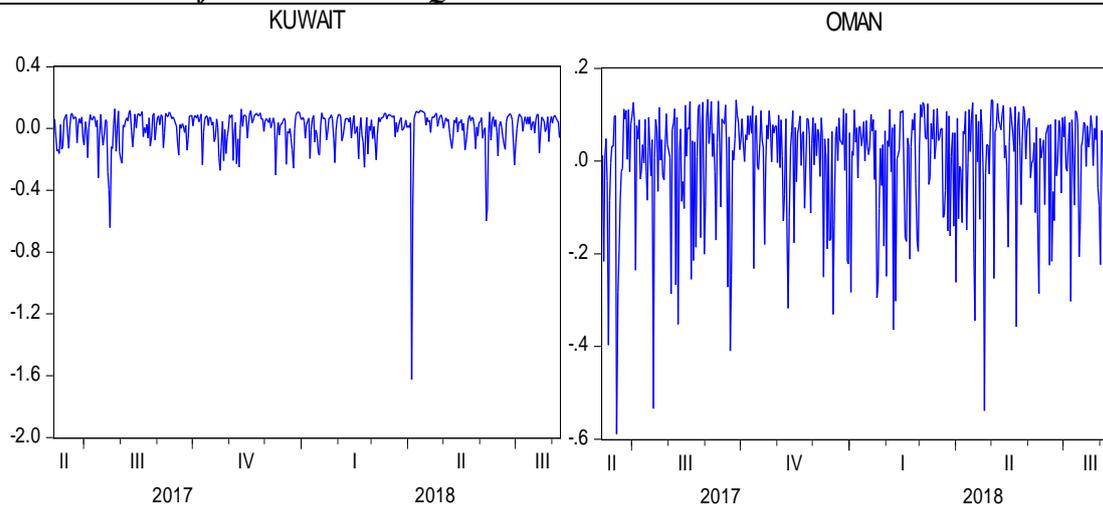

**Table A.3.** Volatility parameters for Kuwait and Oman : Before and after the blockade on Qatar

|  | KUWAIT | OMAN |
|---|---|---|
| *Panel A: Period 1: Before blockade on Qatar* | | |
|  | E-GARCH | T-GARCH |
| Mean equation | | |
| $C$ | 0.169** | 0.0782* |
|  | (0.0013) | (0.0501) |
| Lagged returns | 0.092*** | 0.0452** |
|  | (0.0004) | (0.0010) |
| Variance equation | | |
| $\omega$ | 0.0101*** | 0.0413*** |
|  | (0.0000) | (0.0001) |
| $\alpha$ | -0.0501** | -0.131*** |
|  | (0.0000) | (0.0000) |
| $\beta$ | 0.682*** | 0.719* |
|  | (0.0000) | (0.0351) |
| $\gamma$ | -0.065*** | -0.072*** |
|  | (0.0002) | (0.0000) |
| The duration of persistence: $\alpha + \beta + 0,5\gamma$ | 0.66 | 0.62 |
| The leverage effect: $\gamma$ | -0.065 | -0.072 |
| *Panel B: Period 2: After blockade on Qatar* | | |
|  | T-GARCH | T-GARCH |
| Mean equation | | |
| $C$ | 0.157*** | 0.401*** |
|  | (0.0000) | (0.0000) |
| Lagged returns | 0.121*** | 0.067** |
|  | (0.0003) | (0.0012) |
| Variance equation | | |
| $\omega$ | -0.123*** | -0.115** |
|  | (0.0006) | (0.0023) |
| $\alpha$ | 0.156*** | 0.098*** |
|  | (0.0000) | (0.0004) |
| $\beta$ | 0.502*** | 0.531* |
|  | (0.0008) | (0.0137) |
| $\gamma$ | -0.055** | -0.014*** |
|  | (0.0011) | (0.0007) |
| The duration of persistence: $\alpha + \beta + 0,5\gamma$ | 0.68 | 0.63 |
| The leverage effect: $\gamma$ | -0.055 | -0.014 |

Notes: $\omega$ : the reaction of conditional variance; *α*: the ARCH effect; *β*: the GARCH effect; $\gamma$ : the leverage effect; (.): the p-value; p-value<0.01: ***; p-value<0.05: **; p-value<0.1:*.. With respect to the results of AIC information criterion, we select one lag for all the specifications.

**Fig. A. 2.** Conditional variance of Kuwaiti and Muscat stock returns:
Before and after the blockade on Qatar

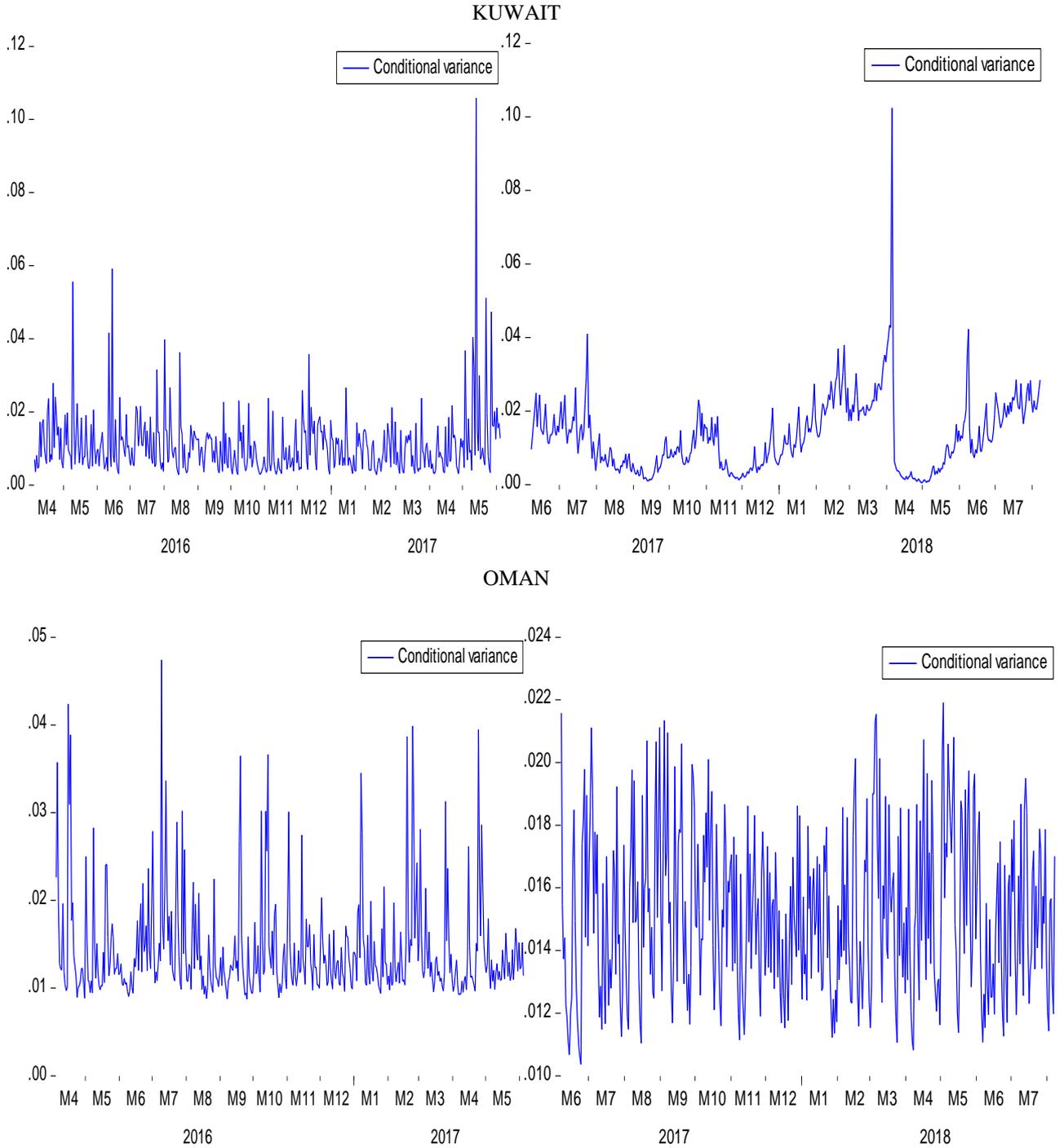

**Table A.4.** Stock market volatility spillovers across Qatar and the boycotting countries (+ Kuwait and Oman): Before and after the blockade on Qatar

| | Qatar | Bahrain | Saudi Arabia | UAE | Egypt | Kuwait | Oman | Contribution from others |
|---|---|---|---|---|---|---|---|---|
| *Panel A. Period 1: Before the blockade on Qatar* | | | | | | | | |
| Qatar | 46.7 | 6.6 | 12.7 | 11.5 | 2.8 | 10.1 | 11.3 | 7.1 |
| Bahrain | 6.5 | 40.2 | 7.9 | 6.3 | 5.3 | 3.6 | 4.1 | 13.8 |
| Saudi Arabia | 19.4 | 7.3 | 57.9 | 13.4 | 5.2 | 9.8 | 6.0 | 5.4 |
| UAE | 8.1 | 5.1 | 8.3 | 46.6 | 4.9 | 9.4 | 3.4 | 4.4 |
| Egypt | 2.2 | 4.2 | 5.3 | 6.5 | 32.3 | 8.1 | 7.4 | 13.1 |
| Kuwait | 21.3 | 7.0 | 8.1 | 5.9 | 6.8 | 42.4 | 9.4 | 11.9 |
| Oman | 19.8 | 4.1 | 4.6 | 4.4 | 4.3 | 8.7 | 39.2 | 12.4 |
| Contribution to others | 22.4 | 7.2 | 24.8 | 22.7 | 4.1 | 4.6 | 2.9 | 68.1 |
| Contribution including own | 69.1 | 47.4 | 82.1 | 69.3 | 36.4 | 47.0 | 48.6 | 49.9 |
| *Panel B. Period 2: After the blockade on Qatar* | | | | | | | | |
| Qatar | 50.0 | 7.3 | 13.4 | 10.7 | 4.3 | 11.3 | 12.4 | 8.3 |
| Bahrain | 7.2 | 44.9 | 8.2 | 7.9 | 6.7 | 4.9 | 5.3 | 14.1 |
| Saudi Arabia | 16.8 | 8.0 | 61.3 | 14.1 | 6.1 | 10.8 | 5.8 | 6.2 |
| UAE | 6.9 | 6.6 | 9.0 | 49.0 | 5.2 | 9.9 | 3.9 | 5.3 |
| Egypt | 3.0 | 5.2 | 6.6 | 7.1 | 39.3 | 9.0 | 6.8 | 13.6 |
| Kuwait | 22.4 | 7.9 | 9.4 | 6.2 | 7.2 | 44.1 | 10.6 | 11.4 |
| Oman | 20.6 | 5.2 | 3.8 | 4.1 | 5.0 | 9.3 | 45.2 | 13.6 |
| Contribution to others | 23.4 | 7.6 | 25.1 | 23.9 | 5.5 | 6.8 | 4.7 | 72.5 |
| Contribution including own | 73.4 | 52.5 | 86.4 | 72.9 | 44.8 | 50.9 | 49.9 | 53.8 |

Notes: The values are calculated from variance decompositions based on 1-step-ahead forecasts. The optimal lag length for the VAR models is 3 for the two periods under study, determined by the Akaike Information Criterion.

**Table A. 5.** The average net directional volatility spillovers across Qatar and the boycotting countries (+ Kuwait and Oman): Before and after the blockade on Qatar

| | Contribution from others | Contribution to others | Average net directional spillover |
|---|---|---|---|
| *Panel A. Period 1: Before the blockade on Qatar* | | | |
| Qatar | 7.1 | 22.4 | 15.6 |
| Bahrain | 13.8 | 7.2 | -6.6 |
| Saudi Arabia | 5.4 | 24.8 | 19.4 |
| UAE | 4.4 | 22.7 | 18.3 |
| Egypt | 13.1 | 4.1 | -9.0 |
| Kuwait | 11.9 | 3.6 | -8.3 |
| Oman | 12.4 | 2.9 | -9.5 |
| *Panel B. Period 2: After the blockade on Qatar* | | | |
| Qatar | 8.3 | 23.4 | 15.1 |
| Bahrain | 14.1 | 7.6 | -6.5 |
| Saudi Arabia | 6.2 | 25.1 | 18.9 |
| UAE | 5.3 | 23.9 | 18.6 |
| Egypt | 13.6 | 5.5 | -8.1 |
| Kuwait | 11.4 | 6.8 | -4.6 |
| Oman | 13.6 | 3.7 | -9.9 |